\documentclass[11pt]{article}
\usepackage{amsmath}
\usepackage{amsfonts}   
 \usepackage{amssymb} 

\numberwithin{equation}{section}

\newcommand{\VS}{{\cal S}}
\newcommand{\symp}{\Omega}
\newcommand{\alg}{{\cal A}}
\newcommand{\sprod}{\mu}
\newcommand{\half}{\frac{1}{2}}
\newcommand{\gr}{\Phi}

\newcommand{\nll}{{\cal N} }
\newcommand{\B}{{\cal B} }
\newcommand{\Q}{{\cal Q}}
\newcommand{\Z}{{\cal Z}}
\newcommand{\C}{C}
\newcommand{\D}{D}

\title{Symmetry Reduction of Quasi-Free States}
\author{C.~G.~Torre \\
{\em Department of Physics} \\ 
{\em Utah State University} \\ 
{\em Logan, Utah, 84322-4415, USA}}

\date{December 2008}

\begin{document}
\maketitle

\begin{abstract}
Given a group-invariant quasi-free state on the algebra of canonical commutation relations (CCR), we show how group averaging techniques can be used to obtain a symmetry reduced CCR algebra and reduced quasi-free state. When the group is compact this method of symmetry reduction leads to standard results which can be obtained using other methods. When the group is non-compact the group averaging prescription relies upon technically favorable conditions which we delineate.  As an example, we consider symmetry reduction of the usual vacuum state for a Klein-Gordon field on Minkowski spacetime by a non-compact subgroup of the Poincar\'e group consisting of a 1-parameter family of boosts, a 1-parameter family of spatial translations and a set of discrete translations.  We show that the symmetry reduced CCR algebra and vacuum state correspond to that used by each of Berger, Husain, and Pierri for the polarized Gowdy ${\bf T}^3$ quantum gravity model.

\end{abstract}

\section{Introduction}

Given a classical field theory on a manifold $M$, if the field equations are invariant under a Lie group $G$ then the equations 
admit a canonical symmetry reduction to a system of equations for fields on $M/G$ defining $G$-invariant solutions of the original field equations \cite{Olver:1993, AFT:2000}.  Other $G$-invariant structures associated with the field theory ({\it e.g.}, Lagrangians, conserved currents) will, under favorable conditions, drop to define corresponding structures for the reduced theory on $M/G$ \cite{Palais:1979,AF:1997, FT:2002}. For non-linear classical field theories symmetry reduction is often the most effective method for obtaining explicit solutions to the field equations and for further analyzing the structure of the field theory. While the theory of symmetry reduction of classical field theories is quite well-understood, in quantum field theory it appears the general theory of symmetry reduction is not nearly so well developed. Nonetheless, quantum models based upon symmetry reduction (usually performed at the classical level) are common in the literature --- particularly literature related to quantum gravity (see, for example, \cite{Bojowald:2008}).   

Motivated by loop quantum gravity models based upon various approaches to quantum symmetry reduction, Engle has given a careful analysis of the axisymmetry reduction of the free Klein-Gordon quantum field theory \cite{Engle:2005}. Using canonical quantization methods, he considers three distinct ways of defining the symmetry reduced quantum theory and exposes the relative merits of each.  He shows that one of these definitions is preferred in the sense that it leads to a complete commutation of the processes of symmetry reduction and canonical quantization as well as possessing other desirable physical features.  

Here we examine symmetry reduction of quantum field theories from the algebraic point of view. We focus on reduction of the bosonic canonical commutation relations (CCR) algebra and the reduction of quasi-free states thereon.  The CCR algebra is determined by a vector space $\VS$ and a symplectic form $\symp$, which can be viewed as representing the phase space of solutions to a set of linear field equations.  The quasi-free states are determined by a scalar product $\sprod$ on $\VS$ which bounds $\symp$. Suppose $G$ is a Lie group acting on $\VS$ preserving $\symp$. Then $G$ defines a group of automorphisms of the CCR algebra. We say that a quasi-free state is $G$-invariant if $\sprod$ is also $G$-invariant.  In this case, one expects $\symp$ and $\sprod$ to somehow induce a reduced symplectic form and a reduced scalar product on the $G$-invariant elements of $\VS$ and hence define a reduced CCR algebra and a reduced quasi-free state.  The principal goal of this paper is to spell out this reduction process in some detail.  

To perform the symmetry reduction we use group averaging techniques, borrowed from  work on constrained quantization and quantum Marsden-Weinstein reduction \cite{Landsman:1993, Giulini:1998}.   Using group averaging, we give a prescription for symmetry reduction of CCR algebras and their quasi-free states.  For reduction by compact groups this prescription always works and gives standard results. In particular,  in the case of the axi-symmetry reduction of the Klein-Gordon field the reduced CCR algebra and reduced vacuum state coincide with that of the usual canonical quantization of the  symmetry reduced classical field theory (as given, {\it e.g.,} in \cite{Engle:2005}).  For non-compact groups the success of the prescription depends upon a number of technical issues which we delineate. These issues largely stem from the fact that, when the symmetry reduction group is non-compact, the group-invariant solutions to the field equations are generally not elements of the space of solutions of the field equations one uses to define $\symp$ and $\mu$.  As an example, we consider a reduction of the massless Klein-Gordon field on Minkowski space by a non-compact subgroup of the Poincar\'e group  consisting of a 1-parameter family of boosts, a 1-parameter family of spatial translations and a set of discrete spatial translations.   We show that, with an appropriate choice for $\VS$, the reduction of the CCR algebra and the vacuum state leads to the algebra and state introduced (apparently independently) by each of Berger, Husain and Pierri (BHP) in their quantization of the polarized Gowdy ${\bf T}^3$ model in general relativity \cite{Berger:1984, Husain:1987, Pierri:2000}.  This result illuminates the fact that dynamical evolution is not unitarily implemented in the Fock representation associated with the BHP vacuum state \cite{Corichi:2002,Torre:2002}.

\section{Quasi-free states on the CCR algebra}

Quasi-free states are defined on the algebra of canonical commutation relations (CCR), a $C^*$ algebra which can be defined as follows.  Fix a symplectic vector space $(\VS, \symp)$, where $\VS$ is a real vector space and $\symp\colon\VS\times\VS\to {\bf R}$ is a symplectic form (possibly degenerate) on $\VS$. Typically, $(\VS, \symp)$ represents the phase space of solutions to some linear field equations. The CCR algebra, denoted $\alg(\VS, \symp)$, is generated by elements denoted $W(\varphi)$, $\varphi\in \VS$, which satisfy
\begin{equation}
W(\varphi)^* = W(-\varphi),\quad W(\varphi_1) W(\varphi_2) = \exp\left\{\frac{i}{2}\symp(\varphi_1,\varphi_2)\right\}
W(\varphi_1+\varphi_2)
\label{Weyl}
\end{equation}
and is completed in a $C^*$ norm, determined as follows (see \cite{Petz:1990} and references therein).
If $\symp$ is non-degenerate the $C^*$ norm is unique, in which case $\alg(\VS,\symp)$ is the Weyl algebra.  If $\symp$ is degenerate, one may introduce the ``minimal regular norm''  to define $\alg(\VS,\symp)$.

 To define a quasi-free state on the CCR algebra we choose an inner product  $\sprod\colon \VS\times\VS\to {\bf R}$ on $\VS$ satisfying for all $\varphi_1,\varphi_2\in \VS$

\begin{equation}
\half|\symp(\varphi_{1},\varphi_{2})	|
\leq
                      \sprod(\varphi_{1},\varphi_{1})^{1/2}
                      \sprod(\varphi_{2},\varphi_{2})^{1/2}.
                      \label{qf}
\end{equation}

Quasi-free states $\omega_\sprod\colon\alg\to{\bf C}$ are in one to one correspondence with such inner products via (see \cite{Petz:1990} and references therein)

\begin{equation}
\omega_\sprod(W(\varphi)) = \exp\left\{-\half \sprod(\varphi,\varphi)\right\}.
\label{state}
\end {equation}

The requirement \eqref{qf} implies that $\symp$ extends to a (possibly degenerate) symplectic form on the real Hilbert space $\VS_\mu$, the completion of $\VS$ with respect to the norm defined by $\sprod$. 
There exists a bounded, skew-adjoint operator $A$ such that

\begin{equation}
\half \symp(\varphi_1,\varphi_2) = \sprod(\varphi_1,A\varphi_2),
\quad \varphi_1,\ \varphi_2\in \VS_\sprod.
\label{adef}
\end{equation}

\section{Symmetry Reduction by Group Averaging}

In this section we shall see that reduction of the CCR and quasi-free states by group averaging always works for compact $G$. For non-compact, unimodular $G$ the success of the group averaging prescription largely depends upon the choice of $\VS$.

We begin with a smooth, faithful group action $\gr\colon G\times \VS \to \VS$, where $G$ is a finite-dimensional Lie group.  We assume the group action preserves the symplectic form and the inner product:
\begin{align}
\symp(\gr_g\cdot\varphi_1,\gr_g\cdot\varphi_2) &= \symp(\varphi_1,\varphi_2),
\label{ginvsymp}\\
\sprod(\gr_g\cdot\varphi_1,\gr_g\cdot\varphi_2) &= \sprod(\varphi_1,\varphi_2),\quad \forall\ g\in G,\quad \varphi_1,\varphi_2\in\VS.
\label{ginvforms}
\end{align}
Thus $G$ acts symplectically and orthogonally on $(\VS,\symp,\sprod)$ and defines a group of $*$-automorphisms of $\alg(\VS,\symp)$ \cite{Petz:1990}.  We remark that the $G$-invariance of $\symp$ and non-degeneracy of $\sprod$ implies that the bounded operators $\gr_g$ and $A$ commute:
\begin{equation}
[A,\Phi_g]=0, \quad \forall\ g\in G.
\label{ainv}
\end{equation}
This follows from
\begin{equation}
\symp(\gr_g\cdot\varphi_1,\gr_g\cdot\varphi_2) = 2\sprod(\varphi_1,\gr_g^\dag A\gr_g\varphi_2)
\end{equation}
and \eqref{ginvsymp}, which imply
\begin{equation}
\sprod(\varphi_1, A\varphi_2 - \gr_g^\dag A\gr_g\varphi_2) = 0,\quad \forall \varphi_1, \varphi_2,
\end{equation}
hence
\begin{equation}
A\gr_g = \gr_g A.
\end{equation}

We formally define averaged bilinear forms via integration with respect to a fixed Haar measure $dg$:
\begin{equation}
\symp_G(\varphi_1,\varphi_2) = \int_G dg\, \symp(\varphi_1,\gr_g\cdot \varphi_2),\quad \sprod_G(\varphi_1,\varphi_2) = \int_G dg\, \sprod(\varphi_1,\gr_g\cdot \varphi_2).
\label{bilinear}
\end{equation}
The integrands are continuous functions of $g\in G$. For compact $G$ these functions are bounded and the integrals exist for all $\varphi_1, \varphi_2\in\VS$. For non-compact $G$ we must assume $\VS$ has been chosen such that the integrals exist for all $\varphi_1, \varphi_2\in\VS$. 

If the measure is bi-invariant, {\it i.e.,} $G$ is unimodular, then the averaged bilinear $\symp_G$ and $\sprod_G$ forms inherit the algebraic symmetries
of $\symp$ and $\sprod$, respectively.  For example,
\begin{align}
\sprod_G(\varphi_2,\varphi_1) &= \int_G dg\, \sprod(\varphi_2,\gr_g\cdot\varphi_1)
\label{sym1}\\
&=\int_G dg\, \sprod(\gr_g\cdot\varphi_1,\varphi_2)\\
&=\int_G dg\, \sprod(\varphi_1,\gr_{g^{-1}}\varphi_2)\\
&=\int_G dg^{-1}\, \sprod(\varphi_1,\gr_{g}\varphi_2)\\
&=\int_G dg\, \sprod(\varphi_1,\gr_{g}\varphi_2)\\
&= \sprod_G(\varphi_1,\varphi_2),
\label{sym2}
\end{align}
where we used the fact that $dg^{-1} = dg$ for a bi-invariant measure.  A similar computation establishes the skew symmetry of $\symp_G$. All compact Lie groups are unimodular. In the non-compact case we henceforth restrict to unimodular groups.\footnote{It should be possible to generalize our considerations to certain non-unimodular groups, but subtleties arise related to the principle of symmetric criticality \cite{Palais:1979,AF:1997, FT:2002} so we leave that investigation for future work.}

The bi-invariance of the measure $dg$ also implies that, for any $h\in G$,
\begin{equation}
\sprod_G(\varphi_1,\gr_h\cdot\varphi_2) = \sprod_G(\varphi_1,\varphi_2),\quad
\symp_G(\varphi_1,\gr_h\cdot\varphi_2) = \symp_G(\varphi_1,\varphi_2).
\label{redinv}
\end{equation}
For example,
\begin{align}
\symp_G(\varphi_1,\gr_h\cdot\varphi_2) &= \int_G dg\, \symp(\varphi_1,\gr_g\gr_h\cdot\varphi_2)\\
&=\int_G dg\, \symp(\varphi_1,\gr_{gh}\cdot \varphi_2)\\
&= \int_G d(gh)\, \symp(\varphi_1,\gr_{gh}\cdot \varphi_2)\\
&=\symp_G(\varphi_1,\varphi_2),
\end{align}
with a similar computation for $\sprod_G$.
Hence the bilinear forms are degenerate: there exist null vectors given by 
\begin{equation}
\varphi = (\Phi_h - 1)\psi,
\label{null}
\end{equation}
for any $h\in G$ and $\psi\in\VS$.
Denote the null spaces of $\sprod_G$ and $\symp_G$  by $\nll_\sprod$ and $\nll_\symp$, respectively.  For the group averaging prescription to work, we require that 
\begin{equation}
\nll_\sprod\subset \nll_\symp,
\label{nullreq}
\end{equation} so that $\sprod_G$ and $\symp_G$ define reduced bilinear forms, respectively $\hat\sprod$ and $\hat\symp$, on $\hat\VS = \VS/\nll_\sprod$.  

For compact $G$, \eqref{nullreq} is always satisfied and, with the Haar measure normalized to unity, the reduced bilinear forms are precisely the restriction of $\symp$ and $\sprod$ to the set $\hat\VS\subset\VS$ of $G$-invariant fields,\footnote{These fields appear in $\VS$ as the trivial representation of $G$.} where $\hat\varphi\in\hat\VS$ satisfies $\gr_g\hat\varphi = \hat\varphi$, $\forall\ g\in G$.   To prove this it is convenient to work on the completion, $\VS_\sprod$, of $\VS$ with respect to the norm defined by $\sprod$. We have the orthogonal decomposition
\begin{equation}
\VS_\sprod = \hat\VS_\sprod \oplus \hat\VS^\perp_\sprod.
\end{equation}
 $\sprod_G$ extends to a bounded bilinear form on $\VS_\sprod$ whence there exists a $\psi\in\VS_\mu$ such that
\begin{equation}
 \sprod_G(\varphi_1,\varphi_2) = \sprod(\varphi_1,\psi).
\end{equation}
From \eqref{redinv} it follows that $\psi$ is $G$-invariant, whence
\begin{equation}\mu_G(\varphi_1,\varphi_2) = \mu_G(\hat\varphi_1,\varphi_2),
\end{equation}
and
\begin{align}
\sprod_G(\varphi_1,\varphi_2) &= \int_G dg\, \mu(\gr_{g^{-1}}\hat\varphi_1,\varphi_2)\\
&=\int_G dg\, \mu(\hat\varphi_1,\varphi_2)\\
&=\mu(\hat\varphi_1,\hat\varphi_2),
\end{align}
where $\hat\varphi_1,\hat\varphi_2$ are the components of $\varphi_1,\varphi_2$ along $\hat\VS_\sprod$.  Evidently, on $\VS_\sprod$ the null space of $\sprod_G$ is precisely $\hat\VS^\perp_\sprod$.  On the quotient $\hat\VS_\sprod = \VS_\sprod/\hat\VS^\perp_\sprod$ the form $\sprod_G$ is non-degenerate. Restricting to $\hat\VS\subset \hat\VS_\sprod$ we obtain the reduced, positive-definite inner product on $\hat\VS$
\begin{equation}
\hat\sprod(\hat\varphi_1,\hat\varphi_2) = \sprod(\hat\varphi_1,\hat\varphi_2).
\end{equation}   
Similarly, on $\VS_\sprod$ the null space of $\symp_G$ includes $\hat\VS_\sprod^\perp$, whence (on $\VS$) $\symp$ drops to define a (possibly degenerate) symplectic form $\hat\symp$ on $\hat\VS$ given by
\begin{equation}
\hat\symp(\hat\varphi_1,\hat\varphi_2) = \symp(\hat\varphi_1,\hat\varphi_2).
\end{equation}
From \eqref{qf} we have
\begin{equation}
\half|\hat\symp(\hat\varphi_{1},\hat\varphi_{2})	|
\leq
                       \left[\hat\sprod(\hat\varphi_{1},\hat\varphi_{1})^{1/2}
                       \right] 
                       \left[\hat\sprod(\hat\varphi_{2},\hat\varphi_{2})^{1/2}
                       \right] ,
                       \label{redqf}
\end{equation}
so that the $\hat\sprod$ defines a quasi-free state $\hat\omega$ on $\alg(\hat\VS,\hat\symp)$.  We say that $\hat \omega$ is the symmetry-reduced state on the reduced algebra  $\alg(\hat\VS,\hat\symp)$.

For non-compact $G$  we cannot  expect any elements of $\VS$ (save the zero vector) to be $G$-invariant (see \S5), and none of the results from the previous paragraph need hold {\it a priori}.  However, under favorable circumstances, a suitable choice of $\VS$ can be found such that the group average converges,  \eqref{nullreq} holds, and we can define $\hat\symp$ and $\hat\sprod$ on $\hat\VS$. If, in addition, $\hat\sprod\geq 0 $ and satisfies \eqref{redqf}, then the group averaging will define a reduced quasi-free state on $\alg(\hat\VS,\hat\symp)$.  Although it appears there is no guarantee that a suitable choice of $\VS$ need exist in general, in \S5 we shall see how the prescription goes through in a non-trivial case. 

Finally, we note that the bi-invariant measure $dg$ is only unique up to rescaling by a positive constant. Hence we actually obtain from the above prescription a 1-parameter family of reduced forms $\hat\symp$ and $\hat\sprod$ and hence  a 1-parameter family of reduced CCR algebras and reduced quasi-free states. However it is straightforward to check that all these algebras are isomorphic with the isomorphisms identifying each of the corresponding quasi-free states.

 \section{Example: Axisymmetric Klein-Gordon field}
 
Here we briefly consider the example studied by Engle  \cite{Engle:2005}, involving a compact symmetry group. The symmetry reduced bilinear forms correspond to the Fock representation of the CCR for the symmetry reduced classical theory, as given in \cite{Engle:2005}.

Let $({\bf M},\eta)$ be $3+1$ Minkowski spacetime with inertial coordinates 
\begin{equation}
x^\alpha = (t,\vec x) = (t,x,y,z),
\label{coord}
\end{equation}
and metric
\begin{equation}
\eta = - dt\otimes dt + dx\otimes dx + dy\otimes dy + dz\otimes dz.
\label{mink}
\end{equation}

We define $\VS$ to be the set of solutions to the Klein-Gordon equation,
\begin{equation}
(\square-m^2)\varphi = -\varphi_{,tt} +  \varphi_{,xx} + \varphi_{,yy} + \varphi_{,zz} - m^2\varphi= 0,
\label{KGeq}
\end{equation}
whose Cauchy data on $t=const.$ hyperplanes are smooth and rapidly decreasing at infinity (elements of Schwarz space). A more explicit representation of elements of $\VS$ is given by
\begin{equation}
\varphi(t,\vec x) = 
\int_{{\bf R}^3} d^3k\, \sqrt{1\over 2\omega(2\pi)^3} \left(a(\vec k) e^{i\vec k\cdot \vec x - i\omega t} + c.c.\right),
\quad \omega= \sqrt{k^2 + m^2},
\label{KGmodes}
\end{equation}
where $a(\vec k)$ is a smooth and rapidly decreasing complex-valued function on ${\bf R}^3$.
The standard Klein-Gordon sumplectic form is given by
\begin{equation}
\symp(\varphi_1,\varphi_2) = \int_\Sigma \sqrt{\gamma}\left(\varphi_2L_n\varphi_1 - \varphi_1 L_n\varphi_2\right) = 
\int_{{\bf R}^3} d^3k\,  i \left(a_1^*(\vec k) a_2(\vec k) - a_1(\vec k) a_2^*(\vec k)\right),
\label{kgsymp}
\end{equation}
where $n$ is the future pointing unit normal to the (arbitrary) spacelike hypersurface $\Sigma$ with induced metric $\gamma$.   

The symmetry group, $G=S^1$, acts via rotations about the $z$ axis:
\begin{equation}
\Phi_\alpha\cdot\varphi(t,x,y,z) = \varphi(t,x\cos\alpha - y\sin\alpha,y\cos\alpha + x\sin\alpha,z).
\end{equation}
As a subgroup of the Poincar\'e group, $G$ clearly acts on $\VS$ and preserves the symplectic form.
The set $\hat\VS\subset \VS$ of $G$-invariant fields is easily seen to be obtained from the solutions $\chi=\chi(t,r,z)$ to
\begin{equation}
-\chi_{,tt} + {1\over r} (r\chi_{,r})_{,r} + \chi_{,zz} - m^2\chi= 0
\end{equation}
on ${\bf M}/G = {\bf R}\times {\bf R}^+\times {\bf R}$
via 
\begin{equation}
\hat\varphi(t,x,y,z) = \chi(t,\sqrt{x^2+y^2},z)
\end{equation}
along with the restriction to smooth solutions with rapidly decreasing Cauchy data. 
The group invariant fields can be expressed as
\begin{equation}
\hat\varphi(t,\vec x) = \int_0^\infty d\kappa\, \int_{-\infty}^\infty dk_z\, \sqrt{\kappa\over4\pi\omega} \left\{A(\kappa,k_z) J_0(\kappa r) e^{i(k_z z -\omega t)} + c.c.\right\},
\label{S1field}
\end{equation}
where $J_0$ is the zeroth order Bessel function,
\begin{equation}
r =\sqrt{x^2 + y^2},\quad \omega = \sqrt{\kappa^2 + k_z^2 + m^2},
\end{equation}
and $A(\kappa,k_z)$ is related to $a(k_x,k_y,k_z)$ via
\begin{equation}
A(\kappa,k_z) =  {1\over 2\pi} \sqrt{\kappa}\int_0^{2\pi} d\beta\,  a(\kappa\cos\beta, \kappa \sin\beta, k_z).
\end{equation}

The usual Poincar\'e invariant Fock vacuum state is defined by 
\begin{equation}
\sprod(\varphi_1,\varphi_2) = \int d^3k\,  \frac{1}{2} \left(a_1^*(\vec k) a_2(\vec k) + a_1(\vec k) a_2^*(\vec k)\right).
\label{kgfock}
\end{equation}
It is easily verified that $\sprod$ is $G$-invariant as it must be since $G$ is a subgroup of the Poincar\'e group.

As described in the \S 3, the symmetry-reduced bilinear forms $\hat\symp$ and $\hat\sprod$  can be computed either by averaging over the group (with normalized measure) or by restricting $\symp$ and $\sprod$ to the $G$-invariant fields \eqref{S1field}. In either case we obtain:
\begin{align}
\hat\symp(\hat\varphi_1,\hat\varphi_2) &=  2\pi i\int_0^\infty d\kappa\, \int_{-\infty}^\infty\, dk_z\left(A_1^*(\kappa,k_z) A_2(\kappa,k_z) - A_1(\kappa,k_z) A_2^*(\kappa,k_z)\right), \\
\hat\sprod(\hat\varphi_1,\hat\varphi_2) &= \pi \int_0^\infty d\kappa\, \int_{-\infty}^\infty\, dk_z\left(A_1^*(\kappa,k_z) A_2(\kappa,k_z) + A_1(\kappa,k_z) A_2^*(\kappa,k_z)\right)
\end{align}
It is straightforward to check that the quasi-free state defined by $\hat\sprod$ on $\alg(\hat\VS, \hat\symp)$ and its associated GNS representation \cite{Petz:1990}  coincide with the usual quantization of the reduced classical field theory as described, {\it e.g.,} in  \cite{Engle:2005}, where $\hat\sprod$ defines the symmetry-reduced Fock vacuum.

\section{Example: BHP vacuum state}

Here we  consider reduction of the massless Klein-Gordon field by a non-compact, disconnected subgroup of the Poincar\'e group.   Besides providing a non-trivial illustration of the prescription described in \S3, this example has a couple of novel features. First, it shows how reduction by a disconnected group can allow one to drop the algebra and quasi-free states to that of a field theory on a compactified manifold.  Second it shows that one may interpret a well-studied state appearing in the quantum gravity ``polarized Gowdy model'' in terms of a reduction of the usual Poincar\'e invariant Fock vacuum for a massless Klein-Gordon field in Minkowski spacetime.  This interpretation provides a simple argument for the lack of unitary implementability of time evolution in  the Fock quantization of the Gowdy model defined by this state.

We begin by describing the group action and characterizing the group-invariant fields.  We start with Minkowski spacetime as described in \eqref{coord}, \eqref{mink}.
Let $M\subset {\bf M}$ be the globally hyperbolic region $t>0$, $t^2 - y^2>0$, equipped with the metric \eqref{mink} restricted to $M$.  Consider the group $G={\bf Z}\times {\bf R}^2$ consisting of (i) translations in $x$ by integral multiples of $2\pi$, (ii) boosts along $y$, (iii) translations in $z$. It is easy to verify that this group acts on $M$. The manifold of orbits, $\hat M = M/G$, is defined by the identifications:
\begin{equation}
x \sim x + 2\pi,\quad \tanh^{-1}({y\over t})\sim  \tanh^{-1}({y\over t}) + {\rm const.},\quad z \sim z + {\rm const.}
\end{equation}
We have $\hat M\approx {\bf R}^+\times {\bf S}^1$.  To see this, consider the diffeomorphism  from $(\tau,\sigma,\eta,\xi)\in {\bf R}^+\times {\bf R}^3$ to $(t,x,y,z)\in M$ defined by
\begin{equation}
t =  \tau\cosh \eta,\quad x =  \sigma,\quad y = \tau\sinh \eta,\quad z = \xi.
\label{diffeo}
\end{equation}
In the  $(\tau,\sigma,\eta,\xi)$ variables the group action on $M$ is defined by $(n,\alpha,\beta)\in {\bf Z}\times {\bf R}^2$:
\begin{equation}
\sigma \to \sigma + 2\pi n,\quad \eta \to \eta + \alpha,\quad \xi\to \xi + \beta,\quad\quad n\in{\bf Z},\quad \alpha,\beta\in{\bf R},
\label{gr}
\end{equation}
whence the quotient space is ${\bf R}^+\times {\bf S}^1$.  The flat metric on $M$ is $G$-invariant and drops to a flat metric $\hat \eta$ on $\hat M$ given by
\begin{equation}
\eta = - d\tau\otimes d\tau + d\sigma\otimes d\sigma,
\end{equation}
where  we identify $\sigma\sim \sigma+2\pi$.
$G$-invariant solutions $\hat\varphi$ of the massless Klein-Gordon equation, \eqref{KGeq} with $m=0$,  are determined by functions $\psi=\psi(\tau,\sigma)$ on $\hat M$ satisfying
\begin{equation}
-\psi_{,\tau\tau} - {1\over \tau} \psi_{,\tau} + \psi_{,\sigma\sigma}  = 0
\label{Gowdyeq}
\end{equation}
via
\begin{equation}
\hat\varphi(t,x,y,z) = \psi(\sqrt{t^2 - y^2}, x).
\label{ginv}
\end{equation}

The reduced field equation \eqref{Gowdyeq} on $\hat M$ is precisely the wave equation satisfied by the scalar field in the polarized Gowdy ${\bf R}\times {\bf T}^3$ spacetime metric by virtue of the symmetry-reduced vacuum Einstein equations  and an appropriate choice of gauge \cite{Berger:1984}.  The symmetry reduced Einstein-Hilbert Lagrangian determines a symplectic form on the space of smooth solutions to \eqref{Gowdyeq} and hence a CCR algebra (see \eqref{C}, \eqref{D}, below). Two representations of this algebra, based upon a pair of algebraic vacuum states, have been studied in some detail in the literature \cite{Berger:1984, Husain:1987, Pierri:2000, Corichi:2002, Torre:2002, Corichi:2006}.  We shall recover one of these states -- the BHP state -- via group averaging.

We define $\VS$ as the set of solutions to the massless Klein-Gordon equation with compactly supported Cauchy data in $M$.    The $G$-invariant symplectic form $\symp$ is given by \eqref{kgsymp}. The resulting CCR algebra $\alg(\VS,\symp)$ is the Weyl algebra of the massless Klein-Gordon field restricted to $M$. The scalar product defining the Poincar\'e invariant Fock vacuum state, \eqref{kgfock}, restricts to $\VS$ to define a $G$-invariant quasi-free state in $\alg(\VS,\symp)$.  There will be no $G$-invariant fields in $\VS$ since fields of the form \eqref{ginv} do not vanish as $|\vec x|\to\infty$.  Nevertheless, $G$ is unimodular and acts on $\VS$ preserving $\symp$ and $\sprod$,  so we may attempt to reduce the CCR algebra and the quasi-free state using group averaging.

It will be convenient to view the bilinear forms $\sprod$ and $\symp$, defined in \eqref{kgsymp} and \eqref{kgfock}, for the set of solutions of \eqref{KGmodes} (with $m=0$), as real and imaginary parts of 
\begin{equation}
\label{ }
\B(\varphi_1,\varphi_2) = i\,\int_{t=0} d^3 x \left(\Phi_1^* \Phi_{2,t} - \Phi_{1,t}^* \Phi_2\right)
\end{equation}
where $\Phi$ is the ``positive frequency'' part of $\varphi$,
\begin{equation}
\varphi(t,\vec x) = 
\int d^3k\, \sqrt{\frac{1}{2\omega(k)(2\pi)^3}} \left(a(\vec k) e^{i\vec k\cdot \vec x - i\omega(k) t} + c.c.\right),
\quad \omega(k)= \sqrt{k^2},
\end{equation}
\begin{equation}
\label{ }
\Phi = 
\int d^3k\, \sqrt{\frac{1}{2\omega(k)(2\pi)^3}} a(\vec k) e^{i\vec k\cdot \vec x - i\omega(k) t},
\end{equation}
and 
\begin{equation}
\symp(\varphi_1,\varphi_2) = -2\, {\rm Im}[\B(\varphi_1,\varphi_2)],
\quad
\sprod(\varphi_1,\varphi_2) =  {\rm Re}[\B(\varphi_1,\varphi_2)] .
\end{equation}
With $g=(n,\alpha,\beta)$, we have
\begin{align*}
\label{}
& \B(\varphi_1,\gr_g\varphi_2)    
=   
\int_{{\bf R}^3} d^3k\, \int_{{\bf R}^3} d^3l\int_{{\bf R}^3} d^3x\, 
  \frac{1}{ 2(2\pi)^3} \sqrt{\frac{1}{ \omega(k)\omega(l)}} a_1^*(\vec l) a_2(\vec k)\\
&\times\left[ \omega(l)+\omega(k)\cosh(\alpha)-k^y \sinh(\alpha)\right]e^{2\pi i k^x n+ i k^z\beta}\\
&\times \exp\Bigg\{i(k^x-l^x) x + i(k^z - l^z) z +iy[\cosh(\alpha) k^y -\sinh(\alpha)\omega(k) - l^y] \Bigg\}.
\end{align*}
The $x,y, z, l^x,l^y, l^z$ integrals are easily performed; after some simplification we have
\begin{equation}
\label{}
\B(\varphi_1,\gr_g\varphi_2)    
=   
-\int_{{\bf R}^3} d^3k\, 
 \sqrt{\frac{1}{ \omega(k)\omega(k,\alpha)}} a_1^*(k^x,l^y(\alpha),k^z) a_2(\vec k)
{dl^y(\alpha)\over d\alpha}
e^{2\pi i k^x n + i k^z\beta},
\end{equation}
where now we define
\begin{equation}
\label{ }
l^y(\alpha) = k^y\cosh(\alpha) - \omega(k)\sinh(\alpha),\quad \omega(k,\alpha) = \sqrt{k^{x2} + l^{y}(\alpha)^{2} + k^{z2}}.
\end{equation}

The group average of $\B$ takes the form
\begin{equation}
\begin{split}
\int_G dg\, \B(\varphi_1,\gr_g\varphi_2)
=
&-\sum_{n=-\infty}^\infty \int_{-\infty}^\infty d\alpha\, \int_{-\infty}^\infty d\beta\,
\int_{{\bf R}^3} d^3k\, 
 \sqrt{\frac{1}{ \omega(k)\omega(k,\alpha)}}\\
 &\times a_1^*(k^x,l^y(\alpha),k^z) a_2(\vec k)
{dl^y(\alpha)\over d\alpha}
e^{2\pi i k^x n + i k^z\beta}.
\label{ga}
\end{split}
\end{equation}
The group average can be evaluated as follows. The sum over $n$ defines the periodic delta function: for a smooth, rapidly decreasing function $f(u)$ we have
\begin{equation}
 \frac{1}{2\pi} \int_{-\infty}^{\infty} du\,\sum_{n=-\infty}^\infty  e^{inu} f(u) = \sum_{n=-\infty}^\infty f(n).
 \end{equation}
 The integral over $\beta$ defines the ordinary delta function: for a smooth, rapidly decreasing function $f(u)$ we have
\begin{equation}
\frac{1}{2\pi} \int_{-\infty}^{\infty} du\, \int_{-\infty}^\infty dv\, e^{ivu} f(u) = f(0).
\end{equation}
We then get
\begin{equation}
\int_G dg\, \B(\varphi_1,\gr_g\varphi_2)
=
2\pi\sum_{n=-\infty}^\infty \int_{-\infty}^\infty dl^y\, 
\int_{-\infty}^\infty dk^y\, 
\frac{ a_1^*(n,l^y,0)}{\sqrt{\omega(n,l^y)}}\frac{ a_2(n, k^y,0)}{\sqrt{\omega(n,k^y)}},
\label{gave}
\end{equation}
where now
\begin{align}
\omega(n,k^y) &= \sqrt{n^2 + k^{y2}},\\
\omega(n,l^y)&=\sqrt{n^2 + l^{y2}},.
\end{align}

We thus have
\begin{equation}
\int_G dg\, \B(\varphi_1,\gr_g\varphi_2)
= \sum_{n=-\infty}^\infty A_{1n}^* A_{2n},
\end{equation}
where we have defined
\begin{equation}
A_n =\frac{\sqrt{2\pi}}{i} \int_{-\infty}^\infty dk\, \frac{1}{(n^2+k^2)^{\frac{1}{4}}}a(n,k,0),\quad n =0, \pm1, \pm 2, \dots
\label{Ap}
\end{equation}
which is a rapidly decreasing sequence of complex numbers.

The null spaces of $\sprod_G$ and $\symp_G$ coincide and consist of the solutions of the Klein-Gordon equation in which $A_{n}=0$, for all $n$.  Taking the quotient by the null space, we see that $\hat\VS$ consists of the set of rapidly decreasing complex sequences, ${\bf A} = \{A_{n},\  n =0, \pm1, \pm2, \dots\}$.
The formula \eqref{Ap} defines the projection $\pi\colon\VS\to\hat\VS$.  
On $\hat\VS$ the reduced bilinear forms are then
\begin{align}
\hat\symp({\bf A}_1, {\bf A}_2) &=  i \sum_{n=-\infty}^\infty \left(A_{1n}^* A_{2n} - A_{1n} A_{2n}^{*}\right)\label{hpsymp}\\
\hat\sprod({\bf A}_1, {\bf A}_2) &= \frac{1}{2} \sum_{n=-\infty}^\infty \left(A_{1n}^* A_{2n} + A_{1n} A_{2n}^{*}\right).
\label{hp}
\end{align}
Both of the reduced forms are non-degenerate and $\hat\sprod>0$.  It is straightforward to check that $\hat\sprod$ bounds $\hat\symp$ and so defines a quasi-free state on $\alg(\hat S,\hat\symp)$.

The bilinear forms \eqref{hpsymp} and \eqref{hp} correspond with those used to define the phase space, CCR algebra and vacuum state in the BHP quantization of the polarized Gowdy ${\bf T}^3$ model \cite{Berger:1984, Husain:1987, Pierri:2000}.  To see this, we first recall the BHP CCR algebra and vacuum state.  

The set of smooth solutions $\Q$ to the reduced field equations \eqref{Gowdyeq} on $\hat M$ take the form
\begin{equation}
\psi(\tau,\sigma) = \frac{1}{\sqrt{4\pi}}a_0(1-i\ln \tau) + \frac{1}{2\sqrt{2}}\sum_{n=-\infty\atop n\ne0}^\infty a_n H_0(|n|\tau) e^{in\sigma} + c.c.,
\label{gowdysol}
\end{equation}
where $a_n$ $n=0,\pm 1,\pm 2,\dots$ is a rapidly decreasing sequence of complex constants. The BHP CCR algebra is defined by $\Q$ and the symplectic form
\begin{equation}
\C(\psi_1,\psi_2) =  i \sum_{n=-\infty}^\infty \left(a_{1n}^* a_{2n} - a_{1n} a_{2n}^{*}\right).
\label{C}
\end{equation}
The BHP vacuum state may be defined by the scalar product
\begin{equation}
\D(\psi_1, \psi_2)= \frac{1}{2} \sum_{n=-\infty}^\infty \left(a_{1n}^* a_{2n} + a_{1n} a_{2n}^{*}\right),
\label{D}
\end{equation}
although the restriction of $\D$ to the zero frequency sector (spanned by $\{a_0,a_0^*\}$) is essentially arbitrary.

Let $\VS_0\subset \VS$ denote the set of massless Klein-Gordon fields which satisfy $a(0,k^y,0)=0$.   Denote by $\hat\VS_0\subset \hat\VS$ the image of $\VS_0$ under the projection $\pi$. $\hat\VS_0$ is the vector space of sequences ${\bf A}_0=\{A_n\}$, $n=\pm1, \pm2, \dots$\ .  In terms of fields on $M$ and $\hat M$ the projection, $\pi\colon\VS_0\to \hat\VS_0$ is given by the group average of the scalar field (see Appendix):
\begin{equation}
\label{grfield}
\pi\varphi(t,\vec x) = \int_G dg\, \gr_g\varphi(t,\vec x) =  \frac{1}{2\sqrt{2}}\sum_{n=-\infty\atop n\neq 0}^\infty \Big(A_n H_0(|n|\tau) e^{inx} + A_n^* H_0^*(|n|\tau) e^{-inx}\Big),
\end{equation}
where $\tau=\sqrt{t^2-y^2}$ and $H_0$ is the zeroth order Hankel function of the second kind.
Indeed, with $\varphi\in \VS_0$ we have 
\begin{equation}
\label{proj}
\int_G dg\, \gr_g\varphi(t,\vec x) = 0 \quad\Longleftrightarrow\quad {\bf A}_0 = 0.
\end{equation}
  Thus we can identify $\hat\VS_0$ with the vector space of solutions $\Q_0\subset \Q$ to the reduced field equations with vanishing zero frequency modes via $a_n=A_n$ and  $a_0=0$.  
With this identification in hand, it is straightforward to verify that the restriction of $\hat\symp$ and $\hat\sprod$ to $\hat\VS_0$ are precisely  the restrictions to $\Q_0$ of the bilinear forms $\C$ and $\D$ used to define the BHP CCR algebra and vacuum state.  

The group averages of elements of $\VS$ do not in general exist as fields on $\hat M$; the putative projection to the zero frequency mode does not converge (see Appendix).\footnote{The group average of an element of $\VS$ can, however, be interpreted as defining a $G$-invariant linear functional on $\VS$ and hence a linear functional on $\hat \VS$.}  Thus, while $\hat S$ is isomorphic to $\Q$, the isomorphism is not determined in the zero mode sector.  Writing $\hat\VS = \Z \oplus \hat\VS_0$, where $\Z$ is the two-dimensional symplectic space of zero frequency solutions, the most general symplectic isomorphism from $\hat\VS$ to $\Q$ is determined by the symplectic isomorphisms of $\Z$. Thus $\hat\mu$ will differ from $\D$ only by the pull back of a symplectic transformation on $\Z$.  

To summarize, aside from zero mode ambiguities, the BHP vacuum state can be interpreted as arising from: (i) construction of a subalgebra $\alg(\VS, \symp)$ of the CCR algebra of the massless Klein-Gordon field by restricting the algebra to solutions with support in $M$; (ii) definition of a quasi-free state $\omega$ on $\alg(\VS, \symp)$ by restriction of the Poincar\'e invariant vacuum state to $\alg(\VS, \symp)$; (iii)  symmetry reduction of $\alg(\VS, \symp)$ and $\omega$ by the group $G$.  

It is known that the GNS representation of the CCR algebra defined by the BHP vacuum state does not allow for unitary implementation of time evolution $\tau\to \tau + const.$  It is now easy to see why this would be the case: the BHP vacuum state comes from symmetry reduction of the  Poincar\'e invariant Klein-Gordon vacuum state which defines a representation of the Klein-Gordon CCR algebra designed to unitarily implement Poincar\'e transformations. Evolution in $\tau$ is not a Poincar\'e 
transformation and, from the results of \cite{Torre:1998}, cannot be expected to be unitarily implemented.

\appendix
\section{APPENDIX: Derivation of \eqref{grfield}}

Here we derive \eqref{grfield}.  The action of  $g=(n,\alpha,\beta)$ on a field $\varphi\in \VS$ is
\begin{align*}
 \gr_g\varphi(t,x,y,z) =   &\int d^3k\, \sqrt{1\over 2\omega(k)(2\pi)^3} \Bigg(a(\vec k) e^{2\pi i k^x n}e^{ik^z \beta} e^{ik^x x  + i k^z z}\\ 
 &\times e^{ i y[k^y\cosh(\alpha) - \omega(k)\sinh(\alpha)]}
 e^{-it[\omega(k)\cosh(\alpha) - k^y\sinh(\alpha)]} + c.c.\Bigg),
 \label{}
\end{align*}
where $\omega(k) = \sqrt{k^{x^2} + k^{y2} + k^{z2}}$.
The group average involves a sum over all $n\in{\bf Z}$ and integrals over $\alpha,\beta\in {\bf R}$. 
The sum over $n$ leads to a periodic delta function in $k^x$ while the integral over $\beta$ yields a delta function in $k^z$. We thus get
\begin{align*}
\label{ }
&\int_G dg\, \gr_g\varphi(t,x,y,z)\\
&=2\pi\sum_{n=-\infty}^\infty\, \int_{-\infty}^\infty d\alpha\, 
\int_{-\infty}^\infty dk^y\, \sqrt{1\over 2\omega(k^y,n)(2\pi)^3} \Bigg(a(n,k^y,0)  e^{in x}\\
&\times e^{ i y[k^y\cosh(\alpha) - \omega(k^y,n)\sinh(\alpha)]}  e^{-it[\omega(k^y,n)\cosh(\alpha) - k^y\sinh(\alpha)]} + c.c.\Bigg),
\end{align*}
 where now $\omega(k^y,n) = \sqrt{n^2 + k^{y2} }$.
 
 Using the coordinates $(\tau,\sigma, \eta, \xi)$ defined in \eqref{diffeo} and changing variables $\alpha\to \alpha + \eta$ we have
 \begin{align*}
&\int_G dg\,  \gr_g\varphi(t,x,y,z)
=2\pi\sum_{n=-\infty}^\infty\, \int_{-\infty}^\infty d\alpha\, 
\int_{-\infty}^\infty dk^y\, \\
&\times \sqrt{1\over 2\omega(k^y,n)(2\pi)^3} \Bigg(a(n,k^y,0)  e^{in \sigma}
e^{ i \tau[k^y\sinh(\alpha) - \omega(k^y,n)\cosh(\alpha)]}   + c.c.\Bigg),
\end{align*}
The $n=0$ term in the above expression is given by
\begin{equation*}
{\cal I} =  \int_{-\infty}^\infty d\alpha\, 
\int_{-\infty}^\infty dk^y\, \sqrt{1\over 2|k^y|(2\pi)^3} \Bigg(a(0,k^y,0) 
e^{ i \tau[k^y\sinh(\alpha) - |k^y|\cosh(\alpha)]}   + c.c.\Bigg),
\end{equation*}
which does not converge for $a(\vec k)$ in Schwarz space.  Henceforth we restrict to elements of $\VS$ which satisfy $a(0,k^y,0)=0$ so that we may restrict to $n\neq 0$ in what follows.

Defining $v$ by
$$
\cosh(v) = {\omega(k^y,n)\over|n|},\quad \sinh{v} = -{k^y\over |n|},
$$
and making the change of variables $\alpha\to \alpha + v$ we have
$$
\int_G dg\, \gr_g\varphi =  2\pi\sum_{n=-\infty\atop n\neq 0}^\infty \int_{-\infty}^\infty d\alpha\, \int dk^y\, \sqrt{1\over 2\omega(k^y,n)(2\pi)^3} \left(a(n,k^y,0) e^{in \sigma - i|n|\tau \cosh(\alpha)} + c.c.\right).
$$
We recall  the integral representation  of the zeroth-order Hankel function of the second kind:
$$
H_0^{(2)}(z) = -{1\over\pi i}\int_{-\infty}^\infty ds\, e^{-iz\cosh s}.
$$
Thus we have
\begin{align*}
\int_G dg\, \gr_g\varphi &=  \frac{\sqrt{\pi}}{2i}\sum_{n=-\infty\atop n\neq 0}^\infty  \int dk\, \sqrt{1\over \omega(k^y,n)} \left(a(n,k,0) e^{in \sigma} H_0^{(2)}(|n|\tau) + c.c.\right)\\
&=\frac{1}{2\sqrt{2}}\sum_{n=-\infty\atop n\neq0}^\infty \left(A_n e^{in\sigma}H_0(|n|\tau)  +  c.c.\right),
\end{align*}
with $A_n$ defined in \eqref{Ap}.

\bibliographystyle{ieeetr}
\bibliography{quasi_free_bibliography}

\begin{thebibliography}{10}

\bibitem{Olver:1993}
P.~Olver, {\em Applications of Lie Groups to Differential Equations}.
\newblock Springer-Verlag, 1993.

\bibitem{AFT:2000}
I.~M. Anderson, M.~E. Fels, and C.~G. Torre, ``{Group Invariant Solutions
  Without Transversality},'' {\em Commun. Math. Phys.}, vol.~212, pp.~653--686,
  2000.

\bibitem{Palais:1979}
R.~Palais, ``The principle of symmetric criticality,'' {\em Communications in
  Mathematical Physics}, vol.~69, pp.~19--30, 1979.

\bibitem{AF:1997}
I.~M. Anderson and M.~E. Fels {\em American Journal of Mathematics}, vol.~119,
  p.~609, 1997.

\bibitem{FT:2002}
M.~E. Fels and C.~G. Torre, ``{The principle of symmetric criticality in
  general relativity},'' {\em Class. Quant. Grav.}, vol.~19, pp.~641--676,
  2002.

\bibitem{Bojowald:2008}
M.~Bojowald, ``Loop quantum cosmology,'' {\em Living Reviews in Relativity},
  vol.~11, no.~4, 2008.

\bibitem{Engle:2005}
J.~Engle, ``{Quantum field theory and its symmetry reduction},'' {\em Class.
  Quant. Grav.}, vol.~23, pp.~2861--2894, 2006.

\bibitem{Landsman:1993}
N.~P. Landsman, ``{Rieffel induction as generalized quantum Marsden-Weinstein
  reduction},'' {\em Journal of Geometry and Physics}, vol.~15, pp.~285--319,
  1995.

\bibitem{Giulini:1998}
D.~Giulini and D.~Marolf, ``{A uniqueness theorem for constraint
  quantization},'' {\em Class. Quant. Grav.}, vol.~16, pp.~2489--2505, 1999.

\bibitem{Berger:1984}
B.~K. Berger, ``{Quantum Effects in the Gowdy $T^3$ Cosmology},'' {\em Annals
  Phys.}, vol.~156, pp.~155--193, 1984.

\bibitem{Husain:1987}
V.~Husain, ``{Quantum Effects on the Singularity of the Gowdy Cosmology},''
  {\em Class. Quant. Grav.}, vol.~4, pp.~1587--1591, 1987.

\bibitem{Pierri:2000}
M.~Pierri, ``{Probing quantum general relativity through exactly soluble
  midi-superspaces. II: Polarized Gowdy models},'' {\em Int. J. Mod. Phys.},
  vol.~D11, p.~135, 2002.

\bibitem{Corichi:2002}
A.~Corichi, J.~Cortez, and H.~Quevedo, ``{On unitary time evolution in Gowdy
  $T^3$ cosmologies},'' {\em Int. J. Mod. Phys.}, vol.~D11, pp.~1451--1468,
  2002.

\bibitem{Torre:2002}
C.~G. Torre, ``{Quantum dynamics of the polarized Gowdy $T^3$ model},'' {\em
  Phys. Rev.}, vol.~D66, p.~084017, 2002.

\bibitem{Petz:1990}
D.~Petz, {\em An Invitation to the Algebra of Canonical Commutation Relations}.
\newblock Leuven University Press, 1990.

\bibitem{Corichi:2006}
A.~Corichi, J.~Cortez, and G.~A. Mena~Marugan, ``{Quantum Gowdy $T^3$ model: A
  unitary description},'' {\em Phys. Rev.}, vol.~D73, p.~084020, 2006.

\bibitem{Torre:1998}
C.~G. Torre and M.~Varadarajan, ``{Functional evolution of free quantum
  fields},'' {\em Class. Quant. Grav.}, vol.~16, pp.~2651--2668, 1999.

\end{thebibliography}
\end{document}